# Title: Switching 2D Magnetic States via Pressure Tuning of Layer Stacking

**Authors:** Tiancheng Song[1†], Zaiyao Fei[1†], Matthew Yankowitz[2], Zhong Lin[1], Qianni Jiang[1], Kyle Hwangbo[1], Qi Zhang[1], Bosong Sun[1], Takashi Taniguchi[3], Kenji Watanabe[3], Michael A. McGuire[4], David Graf[5], Ting Cao[6,7], Jiun-Haw Chu[1], David H. Cobden[1], Cory R. Dean[2], Di Xiao[8] & Xiaodong Xu[1, 7]*

**Affiliations:**
[1]Department of Physics, University of Washington, Seattle, Washington 98195, USA.
[2]Department of Physics, Columbia University, New York, NY, USA
[3]National Institute for Materials Science, Tsukuba, Ibaraki 305-0044, Japan.
[4]Materials Science and Technology Division, Oak Ridge National Laboratory, Oak Ridge, Tennessee 37831, USA.
[5]National High Magnetic Field Laboratory, Tallahassee, FL 32310, USA
[6]Geballe Laboratory for Advanced Materials, Stanford University, CA 94305, USA
[7]Department of Materials Science and Engineering, University of Washington, Seattle, Washington 98195, USA.
[8]Department of Physics, Carnegie Mellon University, Pittsburgh, Pennsylvania 15213, USA.

*Correspondence to: xuxd@uw.edu

†These authors contributed equally to the work.

**Abstract: The physical properties of two-dimensional van der Waals (2D vdW) crystals depend sensitively on the interlayer coupling, which is intimately connected to the stacking arrangement and the interlayer spacing. For example, simply changing the twist angle between graphene layers can induce a variety of correlated electronic phases[1-8], which can be controlled further in a continuous manner by applying hydrostatic pressure to decrease the interlayer spacing[3]. In the recently discovered 2D magnets[9,10], theory suggests that the interlayer exchange coupling strongly depends on layer separation, while the stacking arrangement can even change the sign of the magnetic exchange, thus drastically modifying the ground state[11-15]. Here, we demonstrate pressure tuning of magnetic order in the 2D magnet $CrI_3$. We probe the magnetic states using tunneling[16,17] and scanning magnetic circular dichroism microscopy measurements[10]. We find that the interlayer magnetic coupling can be more than doubled by hydrostatic pressure. In bilayer $CrI_3$, pressure induces a transition from layered antiferromagnetic to ferromagnetic phases. In trilayer $CrI_3$, pressure can create coexisting domains of three phases, one ferromagnetic and two distinct antiferromagnetic. The observed changes in magnetic order can be explained by changes in the stacking arrangement. Such coupling between stacking order and magnetism provides ample opportunities for designer magnetic phases and functionalities.**

## Maintext

In a vdW material, a relative shift of a fraction of a lattice constant between adjacent layers can cause a drastic change in certain physical properties. In particular, if the material is magnetic, it can modify the interlayer exchange pathways leading to a change in the sign of the interlayer exchange coupling[11-15]. For example, bulk $CrI_3$ has monoclinic stacking at room temperature, and undergoes a transition to rhombohedral stacking at 220 K[18]. This stacking has been reported to display ferromagnetic interlayer coupling below the critical temperature of 61 K (Fig. 1a)[18]. On the other hand, thin exfoliated $CrI_3$ has been found to act as a layered antiferromagnetic insulator, in which adjacent ferromagnetic monolayers are antiferromagnetically coupled. Second harmonic generation measurements have revealed a $C_{2h}$ symmetry in bilayer $CrI_3$,[19] consistent with recent theoretical proposals that the antiferromagnetic coupling is associated with monoclinic layer stacking[11-15] (Fig. 1b). Recently, the puncture of a thin flake of $CrI_3$ by a diamond probe tip at low temperature was found to switch the magnetic state from antiferromagnetic to ferromagnetic[20], suggesting that mechanical force can change the layer stacking. These findings highlight the opportunity provided by vdW magnets for realizing new magnetic configurations by controlling the layer stacking arrangement.

Hydrostatic pressure can be used to continuously control the interlayer coupling via the interlayer spacing in vdW crystals. This has recently been shown to modify the bands in graphene/hBN moiré superlattices[21] and transition metal dichalcogenides[22,23], as well as correlated electronic phases in twisted bilayer graphene[3]. Pressure has also been applied to a number of bulk vdW magnets, successfully altering the critical temperature[24-27]. Here, by applying pressure to bilayer and trilayer $CrI_3$, we demonstrate dramatic tuning of the critical field for the spin-flip transition via control of the interlayer spacing, as well as switching of the interlayer magnetic order via a pressure induced structural transition.

Figure 1c shows a schematic of the experimental setup. A magnetic tunnel junction (MTJ) device is composed of a bilayer or trilayer $CrI_3$ sandwiched by top and bottom multilayer graphene contacts. The entire MTJ is encapsulated by hBN to prevent sample degradation. The device is then held in a piston cylinder cell for applying a hydrostatic pressure up to 2.7 GPa. The magnetic states are probed in situ using tunneling measurements. After removal from the cell, reflective magnetic circular dichroism (RMCD) microscopy is performed on the samples with a He-Ne laser. All measurements are at a temperature of 2 K in an out-of-plane magnetic field (see Methods for details).

We first present results from a bilayer $CrI_3$ MTJ (device 1). Figure 1d shows the tunneling current $I_t$ vs magnetic field $H$, swept up and down, at a series of pressures. At zero pressure it shows the typical behavior of a layered antiferromagnetic bilayer[16,17]. Below 0.6 T, the two individually ferromagnetic layers in series form an anti-aligned spin filter, which suppresses the tunneling current to form a plateau. As the field is increased a spin-flip transition occurs to a fully polarized state with a higher tunneling current.

As the pressure is increased the critical field for the spin-flip transition rises dramatically to above 1.3 T (Fig. 1d & Fig. 2a), more than double its zero-pressure value. Such an enhancement can be explained by reduced interlayer spacing, which increases the wavefunction overlap and thus the interlayer exchange strength[11-15]. $I_t$ also increases substantially with pressure owing to the reducing interlayer spacing. In contrast, the critical temperature $T_c$ only increases slightly (Fig. 2a and Supplementary Fig. 1), consistent with it being mainly determined by intralayer exchange interactions which are relatively independent of interlayer spacing. Note that the steady background decrease of $I_t$ with magnetic field is due to the positive magnetoresistance in the multilayer graphene contacts[28]. This background magnetoresistance becomes more noticeable as the tunneling magnetoresistance of the $CI_3$ is reduced, i.e., at higher pressures.

Importantly, at the highest pressure of 2.7 GPa, the jump in $I_t$ vs $H$ due to the spin-flip transition is absent, and only the background multilayer graphene magnetoresistance remains. Figure 2b compares the temperature dependence of the low-bias tunneling conductance at $\mu_o H$=0 prior to applying pressure (black) with that at 2.7 GPa (red). Initially at zero pressure there is a kink near 44 K, consistent with $T_c$ reported for the layered antiferromagnet bilayer[10]. Below $T_c$, $I_t$ decreases on cooling due to strengthening antiferromagnetic order. In contrast, at 2.7 GPa $I_t$ increases on cooling, indicative of ferromagnetic order. In fact, this temperature dependence is similar to that of the fully spin-polarized state with $\mu_o H$=1.5 T applied at zero pressure (Supplementary Fig. 1a). These observations taken together imply that the bilayer $CrI_3$ has switched from a layered antiferromagnetic state at low pressure to a ferromagnetic state at high pressure.

To further support this conclusion, we performed RMCD measurements after the devices were removed from the pressure cell. Device 1 was found to be broken, so the RMCD was done on a second bilayer device (device 2) which had been cycled to a comparable pressure (2.45 GPa) and temperature. Figure 2c shows the ambient-pressure RMCD signal at 2 K, which exhibits a single pronounced hysteresis loop centered at $\mu_o H$=0, characteristic of ferromagnetism (see Supplementary Fig. 2 for the full data set). This is distinct from the vanishing RMCD signal from a pristine $CrI_3$ bilayer (Fig. 2d)[10]. Figure 2e shows the Raman spectra of a pristine bilayer (black trace) and of device 2 after pressure (red trace), taken at 80 K. Compared to the pristine samples, the peaks near 77 $cm^{-1}$ and 105 $cm^{-1}$ are blue and red shifted by about 0.8 $cm^{-1}$ and 0.6 $cm^{-1}$, respectively. This is consistent with Raman studies of $CrI_3$ bulk crystals, in which these peaks shift in a similar manner by about 1 $cm^{-1}$ as a structural transition occurs from rhombohedral to monoclinic[29]. Since these different stacking arrangements of $CrI_3$ have opposite signs of the interlayer exchange, we infer that the high pressure induces an irreversible structural transition in bilayer $CrI_3$.

Having demonstrated pressure control of magnetic order in bilayers, we now consider trilayer $CrI_3$. A pristine exfoliated trilayer has two layered antiferromagnetic ground states, which we will label ↑↓↑ and ↓↑↓ with obvious notation. Figure 3a shows $I_t$ vs $H$ at zero pressure for a trilayer MTJ device (device 3). Either ground state presents two anti-aligned spin filters in series, which as in the bilayer case produces a low-current plateau at low field. A spin-flip transition occurs at ~1.6 T to a fully polarized state (↑↑↑ or ↓↓↓), causing a sudden jump to a high-current plateau at higher field. At a moderate applied pressure of 1.2 GPa (Fig. 3b), the $I_t$-$H$ traces have the same form as those at zero pressure but the critical field for the spin-flip transition is increased. At the highest pressure of 2.45 GPa (Fig. 3c), this critical field reaches as high as 3.7 T, more than double the value at zero pressure (see inset to Fig. 3b, and Supplementary Fig. 3).

At high pressure, however, additional features appear which have not previously been seen in trilayer $CrI_3$. Figure 3c shows initial $I_t$-$H$ traces at 2.45 GPa. In addition to the usual low- and high-field current levels, we observe a new intermediate level, suggesting another degree of freedom is involved. After performing these measurements, we swept the DC bias while monitoring $I_t$ at a fixed magnetic field of +1.3 T, as shown in the inset to Fig. 3c. While first increasing the bias (red trace), at about $V$=-300 mV we observed a sudden jump of $I_t$ to a higher level. Thereafter, the current appeared to remain in the higher level as the bias was returned to zero (blue trace), indicating a permanent change had occurred in the magnetic configuration. Such a permanent change could be caused by a change in crystal structure, such as a reconfiguration of the stacking. Figure 3d shows an $I_t$-$H$ trace after this reconfiguration occurred. The lower field jump is at the same position as before (1.7 T, Fig. 3c), but the higher field jump has disappeared, and the current at low field roughly doubled. These observations can be explained naturally as follows.

Prior to the current-induced reconfiguration (Fig. 3c), in the low-field current plateau the sample contains coexisting domains of two distinct layered antiferromagnetic phases, which we will call AFM I and II. AFM I is the phase found in the pristine trilayer, with two antiferromagnetic interfaces and hence one time-reversal pair of ground states, {↑↓↑, ↓↑↓}. AFM II is a new phase having one antiferromagnetic and one ferromagnetic interface, with two time-reversal pairs of possible magnetic configurations, {↑↑↓, ↓↓↑} and { ↓↑↑, ↑↓↓}. This situation could occur as a result of a change in the stacking at just one of the two interlayer interfaces, with similar consequences as for the single interface in the bilayer. The case with the upper interface ferromagnetic and the lower antiferromagnetic is sketched as one of the insets in Fig. 3c. In either AFM I or AFM II phase, as the field is increased at some point the Zeeman energy overcomes the antiferromagnetic coupling to produce a spin-flip transition. Since AFM II has only one antiferromagnetic interface, the AFM II domains flip to the fully polarized configuration at a lower field of 1.7 T, resulting in the intermediate current level. Since AFM I has two antiferromagnetic interfaces, the AFM I domains switch at roughly double the field, 3.7 T, to leave the sample fully polarized and giving the highest current level. After the current-induced reconfiguration (Fig. 3d), the disappearance of only the 3.7 T jump implies that most of the AFM I has converted to AFM II. Closer inspection of Fig. 3d nevertheless reveals a tiny hysteresis remaining up to the higher critical field, indicating that a small amount of AFM I is still present.

After removing device 3 from the pressure cell we performed spatially resolved RMCD measurements that confirm the presence of multiple magnetic phases. Figure 4a shows a spatial map of the RMCD signal at a fixed field of 2 T, which is sufficient to fully polarize the $CrI_3$. (Note that in addition to trilayer we also see bilayer and multilayer regions in this sample; see Supplementary Fig. 4). Figures 4b-d show the RMCD signal as a function of magnetic field at the three indicated laser spot positions. At position P (Fig. 4b), the RMCD exhibits the three-flip AFM I behavior of a pristine exfoliated trilayer. At position Q (Fig. 4c), near the tunnel junction area, there are again three spin-flip transitions, but they occur at fields about half those for the pristine trilayer. This suggests the magnetic state at Q is AFM II, consistent with the assignment from the tunneling measurements. At position R (Fig. 4d), the RMCD signal shows a single hysteresis loop centered at zero field, characteristic of a ferromagnetic phase in which both interfaces in the trilayer have ferromagnetic coupling.

The distinct spin-flip transition fields can be used to uniquely identify domains of different phases. For example, when the field is swept down from 2.0 to 1.3 T, only AFM I has a spin-flip transition, with a change in RMCD signal of about 1.3%. A map of AFM I domains can therefore be created by mapping the RMCD signal at these two fields and identifying the regions where the signal is different in the two cases (inset in Fig. 4b). A similar procedure yields maps of the AFM II and fully ferromagnetic domains (insets in Fig. 4c and 4d, respectively; details are given in Supplementary Fig. 5).

The existence of three distinct magnetic phases implies that at least three different stacking configurations in trilayer $CrI_3$ can be accessed by applying pressure. As in the bilayer case, Raman spectroscopy also shows variations consistent with multiple stacking types (Supplementary Fig. 4). The coexistence of the different stacking configurations is likely to involve inhomogeneity in the sample which causes the conditions for the first-order transition to vary from point to point. In future work, atomically resolved imaging (e.g. transmission electron miscopy) could enable identification of exact stacking configurations, details of domain boundaries, and their precise connections with the various magnetic states. Our work highlights $CrI_3$ as a model system for exploring reconfigurable magnetic structure via control of stacking, by pressure, or by van der Waals assembly, potentially with the addition of twist-angle control. This suggests new possibilities for engineering magnetism, such as creating designer real-space spin textures[30], and manipulating magnetic order to control electronic phenomena, for example in a layered Chern insulator such as $MnBi_2Te_4$.[31,32]

**Acknowledgments:** The authors acknowledge Shiwei Wu and Weida Wu for the insightful discussion. This work was mainly supported by the Department of Energy, Basic Energy Sciences, Materials Sciences and Engineering Division, Pro-QM EFRC (DE-SC0019443). The device fabrication and quantum tunneling measurement are partially supported by NSF MRSEC 1719797, and the magnetic circular dichroism measurement is partially supported by the Department of Energy, Basic Energy Sciences, Materials Sciences and Engineering Division (DE-SC0018171). The material synthesis at UW is partially supported by the Gordon and Betty Moore Foundation's EPiQS Initiative, Grant GBMF6759 to JHC. MAM was supported by the US Department of Energy, Office of Science, Basic Energy Sciences, Materials Sciences and Engineering Division. KW and TT acknowledge support from the Elemental Strategy Initiative conducted by the MEXT, Japan, A3 Foresight by JSPS and the CREST (JPMJCR15F3), JST. XX and JHC acknowledge the support from the State of Washington funded Clean Energy Institute. XX also acknowledges the support from the Boeing Distinguished Professorship in Physics.

**Author Contributions:** XX, TS, MY, CD and DX conceived the experiment. TS and ZF fabricated and characterized the devices, assisted by MY and BS. TS, ZF and MY performed the high-pressure measurements, assisted by DG. TS performed magnetic circular dichroism and Raman measurements. KH and QZ assisted in Raman measurement. TS, ZF, XX, DX, TC, DC analyzed and interpreted the results. MM, ZL, QJ and JHC independently synthesized and characterized the bulk $CrI_3$ crystals. TS, XX, DC, DX, ZF wrote the paper with inputs from all authors. All authors discussed the results.

**Author Information:** The authors declare no competing financial interests. Correspondence and requests for materials should be addressed to xuxd@uw.edu.

## Main Figures

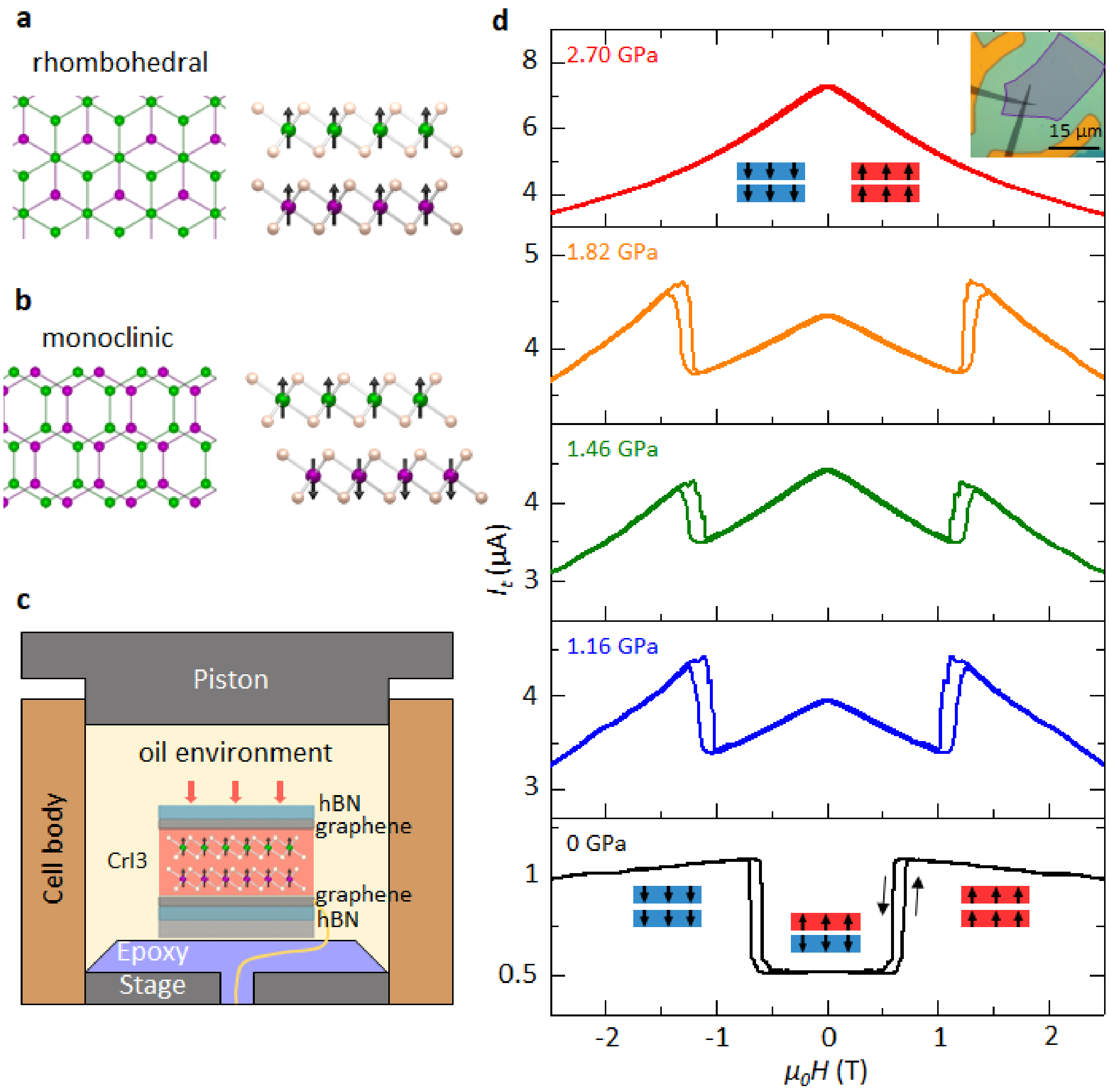

**Figure 1 | Stacking-order determined 2D magnetism and tunneling measurements of bilayer $CrI_3$ under pressure. a**, Schematic of rhombohedral stacking with top (left) and side view (right) indicating the ferromagnetic interlayer coupling. **b**, Same for monoclinic stacking, indicating the antiferromagnetic interlayer coupling. **c**, Schematic of high-pressure experimental setup. A $CrI_3$ magnetic tunnel junction (MTJ) device is held in a piston pressure cell. **d**, Tunneling current $I_t$ vs magnetic field $H$ at a series of pressures. Insets: magnetic states (only one of the two time-reversal AFM ground states is shown) and optical microscope image of a bilayer MTJ (device 1) with junction area smaller than 1 μm$^2$. Applied DC bias: 50 mV.

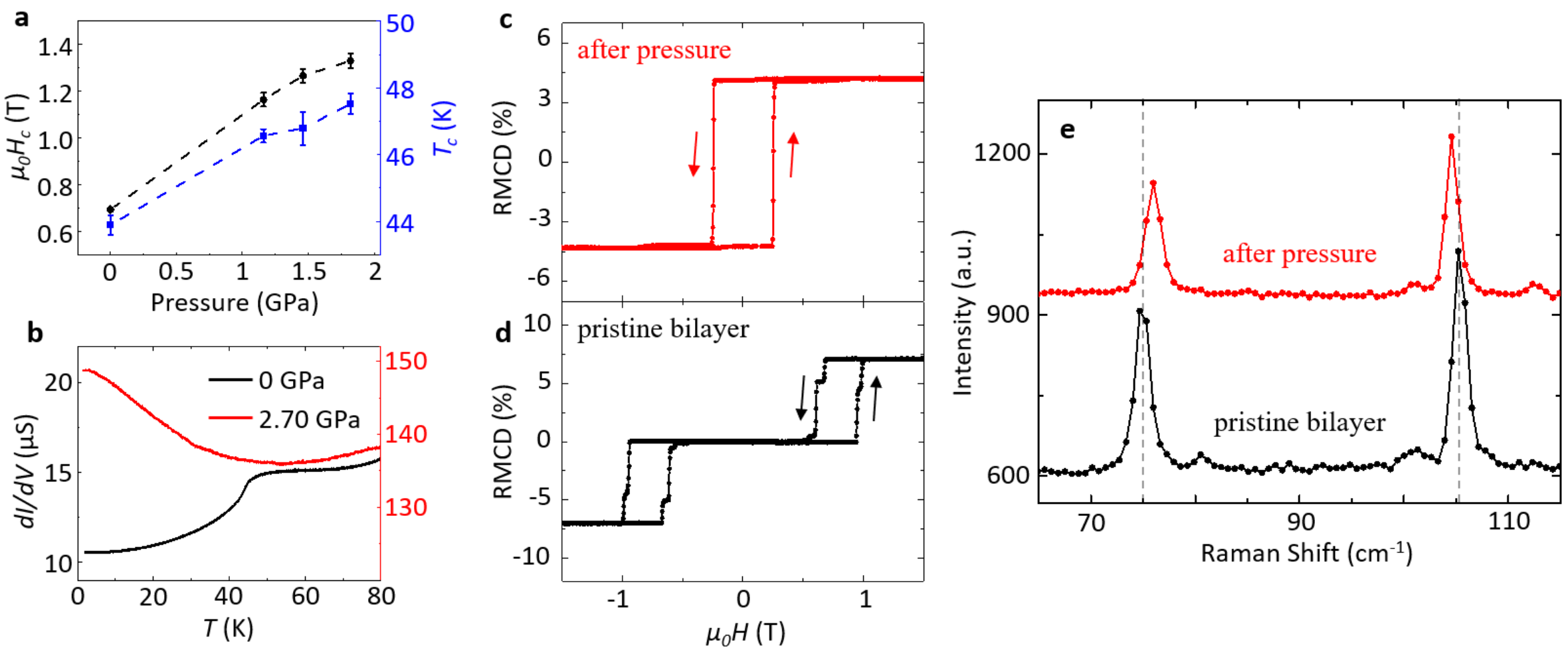

**Figure 2 | Effect of pressure on the magnetic properties of bilayer $CrI_3$. a**, Extracted critical field for the spin-flip transition (black filled circles) and critical temperature (blue squares) as a function of pressure. The error bar for the critical field is determined by the half width of the spin-flip transition. For the critical temperature, the error bar is determined by the temperature range over which dG/dT drops to 80% of the peak value. **b**, Low-bias tunneling conductance vs temperature at zero pressure (black) and 2.7 GPa (red). **c,** Reflective magnetic circular dichroism (RMCD) signal from another bilayer (device 2) after removal from pressure cell where it underwent a comparable pressure (2.45 GPa) and thermal cycle to device 1, and **d**, from a pristine bilayer $CrI_3$. **e**, Raman spectra of bilayer device 2 after pressure (red trace) and of a pristine bilayer (black trace). Data are vertically shifted for clarity.

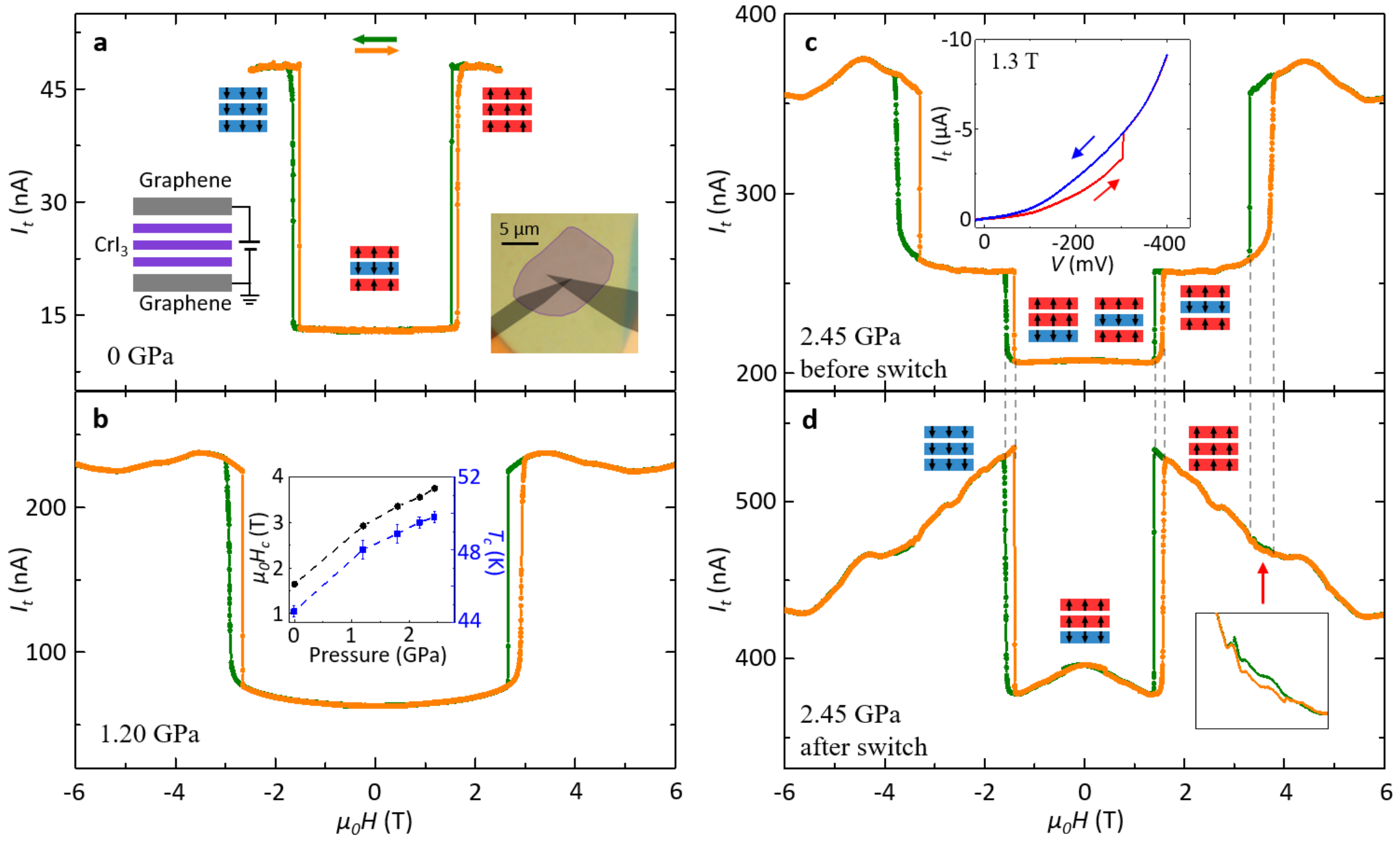

**Figure 3 | New pressure-induced magnetic states in trilayer $CrI_3$.** Tunneling current as a function of magnetic field at **a**, zero pressure and **b**, 1.2 GPa. Applied bias: 70 mV. Inset in (a) are schematics indicating the magnetic states, the measurement geometry (bottom left), and an optical microscope image of device 3 (bottom right). The inset in (b) shows the extracted critical field for the spin-flip transition (black filled circles) and critical temperature (blue squares) as a function of pressure. **c**, Tunneling current vs magnetic field at 2.45 GPa. Inset: tunneling current measured while sweeping bias up (red) and down (blue). The sudden jump implies an irreversible change in the sample configuration. Two magnetic phases, AFM I and AFM II, coexist at low fields; see text for details. **d**, Tunneling current vs magnetic field measured after this switching event. Insets show only one of the two possible AFM II magnetic configurations for the low-current level, as mentioned in the text. Also inset is a zoom on the region indicated. The background quantum oscillations originate from the multilayer graphene contacts.

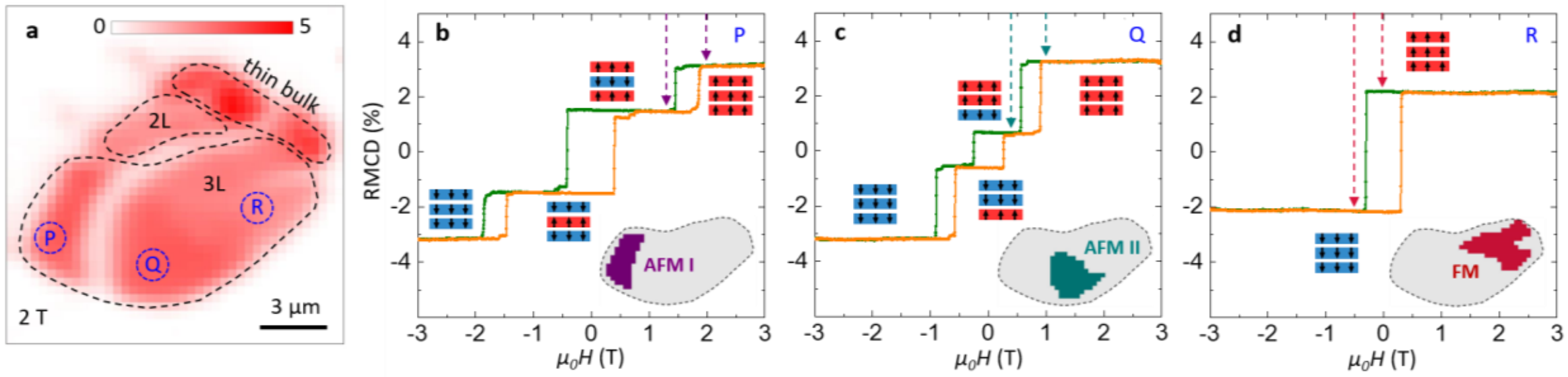

**Figure 4 | Magnetic and layer stacking phase map of a trilayer $CrI_3$ flake. a**, Spatial map of RMCD signal at 2 T after sample removal from the pressure cell. Regions of different thickness are labelled. The low-signal curved band corresponds to a crack in the sample. **b-d**, RMCD signal vs magnetic field at the spots P, Q and R indicated in (a), respectively. Inferred magnetic states are shown in the insets. Insets also show the spatial maps identifying domains of each phase, in each case using the difference in RMCD signal between the two fields on the above trace. See Supplementary Fig. 5 for details. AFM I: both interfaces antiferromagnetically coupled. AFM II: one interface antiferromagnetically coupled and the other ferromagnetically coupled. FM: both interfaces ferromagnetically coupled. For AFM II phase, there are two pairs of possible magnetic configurations: {↑↑↓, ↓↓↑} and { ↓↑↑, ↑↓↓ }, only the first of which are illustrated in (c).

**Methods:**

**Device fabrication:** The multilayer-layer graphene and hBN flakes of 20-40 nm were mechanically exfoliated onto either 285 nm or 90 nm $SiO_2$/Si substrates and examined by optical and atomic force microscopy under ambient conditions. Only atomically clean and smooth flakes were used for making devices. V/Au (5/50 nm) metal electrodes were deposited onto the bottom hBN flakes and substrates using a standard electron beam lithography with a bilayer resist (A4 495 and A4 950 poly (methyl methacrylate (PMMA))) and electron beam evaporation. $CrI_3$ crystals were exfoliated onto 90 nm $SiO_2$/Si substrates in an inert glove box with water and oxygen concentration less than 0.5 ppm. The $CrI_3$ flakes were identified by their optical contrast relative to the substrate using the established optical contrast models of $CrI_3$[10]. The layer assembly was performed in the glove box using a polymer-based dry transfer technique. The flakes were picked up sequentially: top hBN, top graphene contact, $CrI_3$, bottom graphene contact. The resulting stacks were then transferred and released on top of the bottom hBN with pre-patterned electrodes. In the resulting heterostructure, the $CrI_3$ flake is fully encapsulated, and the top/bottom graphene flakes are connected to the pre-patterned electrodes. Finally, the polymer was dissolved in chloroform for less than one minute to minimize the exposure to ambient conditions. The $SiO_2$/Si substrates were diced to approximately 1.7 mm by 1.7 mm to fit into the inner bore of the pressure cell[21].

**Electrical measurement with pressure control:** The electrical measurements at zero applied pressure were performed in a PPMS DynaCool cryostat (Quantum Design, Inc.) with a base temperature of 1.7 K. The measurements at high pressure were performed with a piston pressure cell in a VTI insert cryostat under similar experimental conditions. Figure 1c shows the schematic of experimental setup. For DC measurement, a bias voltage ($V$) is applied to the top graphene contact with the bottom one grounded. The resulting tunneling current ($I_t$) is amplified and measured by a current preamplifier (DL Instruments; Model 1211). For the AC measurement, a standard lock-in technique is used by applying a 500 µV AC excitation at a relatively low frequency of about 13 Hz with Stanford Research Systems SR830.

The hydrostatic pressure was applied using a pressure cell. The device was first glued to a metal stage using epoxy, then Pt wires were affixed to the gold contacts using silver paste. A Teflon cup was filled with the pressure medium (oil) and carefully fitted over the device and onto the stage, such that the device is completely encapsulated in oil. The stage/Teflon cup was then fitted into the inner bore of a piston cylinder cell and a hydraulic press was used to compress the top of the Teflon cup which was held in place by a locking nut. The pressure cell was then loaded into a cryostat for electrical measurement. The in-situ pressure was determined by measuring the fluorescence response of a ruby crystal in the cell through a thin optical fiber at both room and low temperature. Increasing or decreasing pressure requires warming the sample to room temperature and re-loading the cell in the hydraulic press before cooling again. This technique follows closely to the previous study on van der Waals heterostructures with pressure[21].

**Reflective magnetic circular dichroism and Raman spectroscopy measurements:** The reflective magnetic circular dichroism (RMCD) measurements were performed in a closed-cycle cryostat (attoDRY 2100) at a temperature of 2 K and an out-of-plane magnetic field up to 9 T. A 632.8 nm HeNe laser was used to probe the device at normal incidence with a fixed power of 100 nW. The AC lock-in measurement technique used to measure the RMCD signal follows closely to the previous magneto-optical Kerr effect (MOKE) and RMCD measurements of the magnetic order in atomically-thin $CrI_3$[10,16]. The collinear-polarization low-frequency Raman spectroscopy was performed with 150 μW 632.8 nm HeNe laser at a temperature of 80 K.

**Data Availability:** The data that support the findings of this study are available from the corresponding author upon reasonable request.

# Supplementary Information for

# Switching 2D Magnetic States via Pressure Tuning of Layer Stacking

**Authors:** Tiancheng Song[1†], Zaiyao Fei[1†], Matthew Yankowitz[2], Zhong Lin[1], Qianni Jiang[1], Kyle Hwangbo[1], Qi Zhang[1], Bosong Sun[1], Takashi Taniguchi[3], Kenji Watanabe[3], Michael A. McGuire[4], David Graf[5], Ting Cao[6,7], Jiun-Haw Chu[1], David H. Cobden[1], Cory R. Dean[2], Di Xiao[8] & Xiaodong Xu[1, 7]*

[*]Correspondence to: xuxd@uw.edu

†These authors contributed equally to the work.

**This file contains supplementary figures 1-5.**

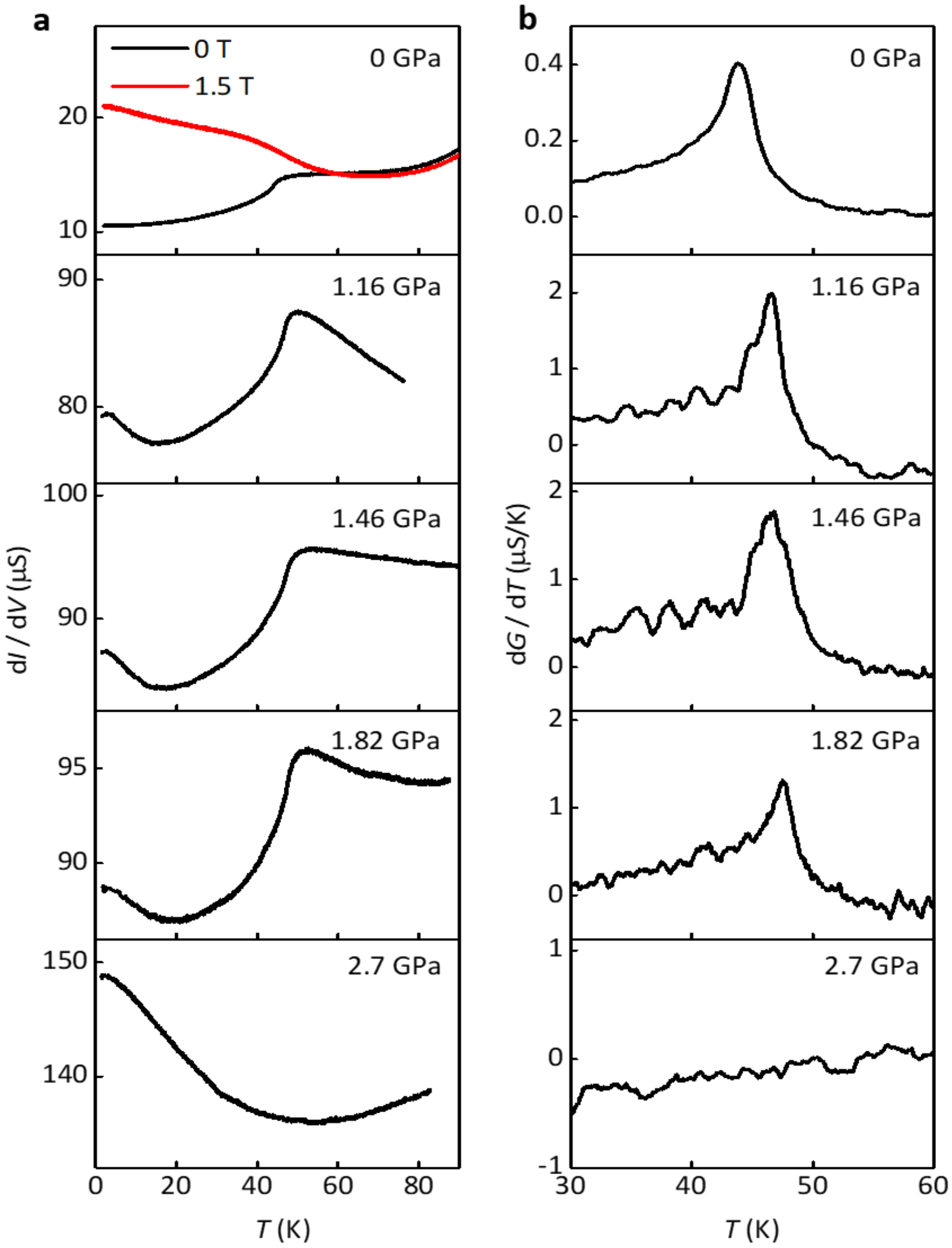

**Supplementary Fig. 1 | Zero bias tunneling conductance vs temperature of bilayer MTJ device 1 at a series of pressures.** From top to bottom, pressure is set at 0, 1.16 GPa, 1.46 GPa, 1.82 GPa and 2.7 GPa respectively. **a**, zero bias tunneling conductance, and **b**, its derivative with respect to temperature vs temperature. The peaks of the traces in (b) correspond to critical temperature. At zero pressure, bilayer is in fully spin polarized state at a magnetic field of 1.5 T. The tunneling conductance increases as temperature decreases (red trace, top panel in (a)), which is similar to that at 2.7 GPa (bottom panel in (a)), indicative of ferromagnetic phase.

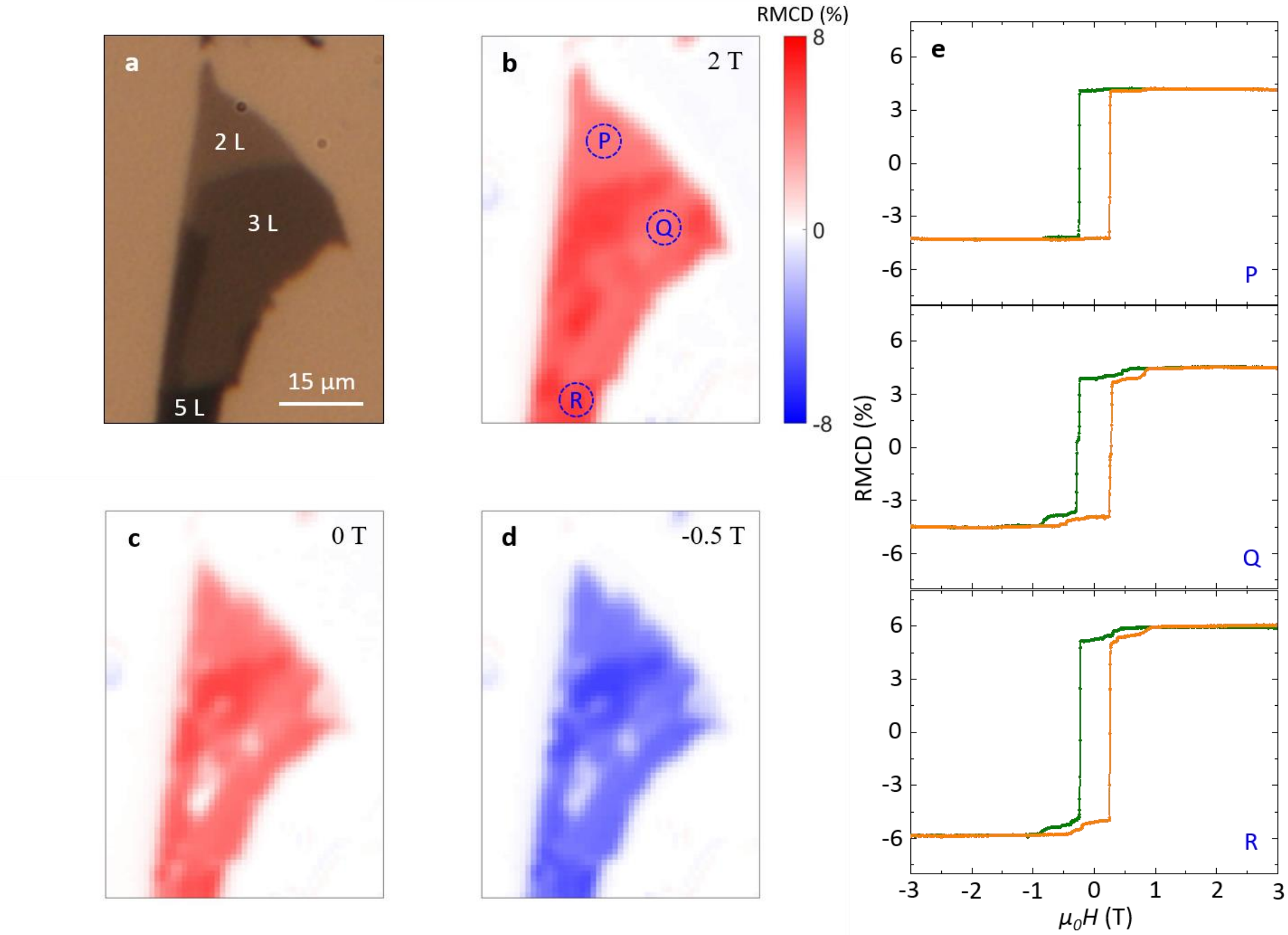

**Supplementary Fig. 2 | Reflective magnetic circular dichroism (RMCD) measurements of a bilayer $CrI_3$ device 2 after removing from the pressure cell. a**, Optical microscope image of the $CrI_3$ flake before device fabrication. Layer thickness is indicated. **b-d**, RMCD signal maps at magnetic field of 2, 0 and -0.5 T, respectively. **e**, RMCD signal vs magnetic field *H* at selected spatial points indicated in (b). Hysteresis centered at zero field is observed for all three points, characteristic of ferromagnetism.

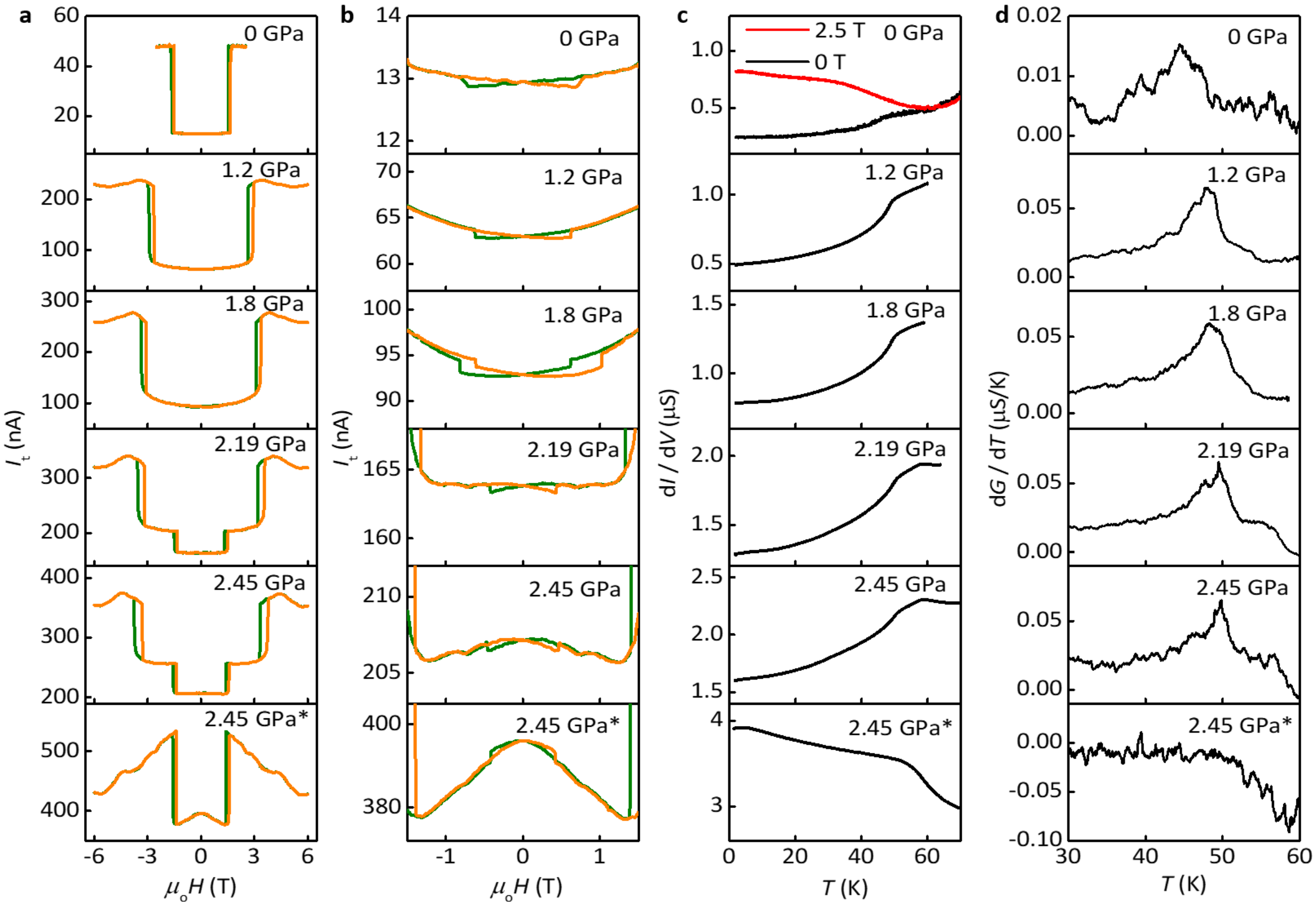

**Supplementary Fig. 3 | Additional tunneling measurements for trilayer $CrI_3$ (device 3). a**, Tunneling current vs magnetic field at a series of pressures, and **b**, zoom in plots of the low current level. Pressure from top to bottom: 0, 1.2 GPa, 1.8 GPa, 2.19 GPa and 2.45 GPa before and after magnetic state switching. Upon close inspection, figure b shows two small but sharp jumps between two discrete values of $I_t$ within the low-current region. We associate these jumps with the switching between two states, ↑↓↑ and ↓↑↓. These states exhibit different values of $I_t$ owing to their opposite magnetic moments, which results in different tunnel barriers due to the Zeeman shifts arising at finite magnetic field. **c**, Zero bias tunneling conductance vs temperature at a series of pressures. At zero pressure and magnetic field of 2.5 T, trilayer is in fully spin polarized. The zero bias tunneling conductance increases as temperature decreases (red trace, top panel), similar to that at 2.45 GPa after magnetic state switching (bottom panel). **d**, Zero bias tunneling conductance's derivative with respect to temperature vs temperature.

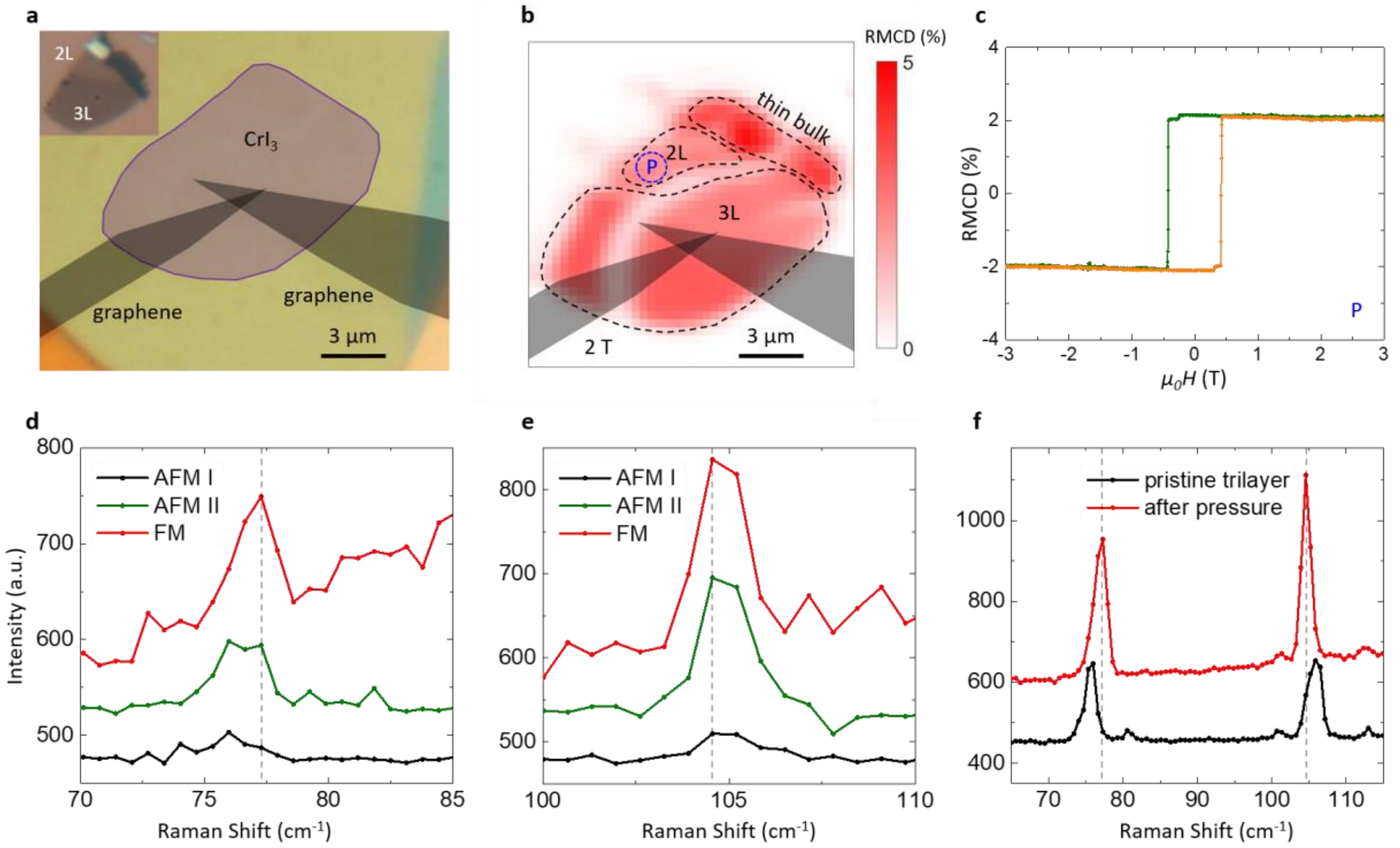

**Supplementary Fig. 4 | Additional optical measurements of trilayer $CrI_3$ devices. a**, False color optical microscope image of the trilayer MTJ device 3. Inset: optical microscope image of the trilayer flake with indicated layer thickness before device fabrication. The flake developed cracks during heterostructure fabrication. **b**, RMCD signal map at a magnetic field of 2 T. **c**, RMCD signal vs magnetic field of the bilayer flake on the top left corner of this device, showing ferromagnetism after pressure, consistent with other bilayer devices. **d & e**, Raman spectra of three different magnetic phases, AFM I (black), AFM II (green), and FM (red) in trilayer. **f**, Comparison of Raman spectra of the trilayer in device 2 after pressure with a pristine trilayer, implying pressure induced structural phase transition.

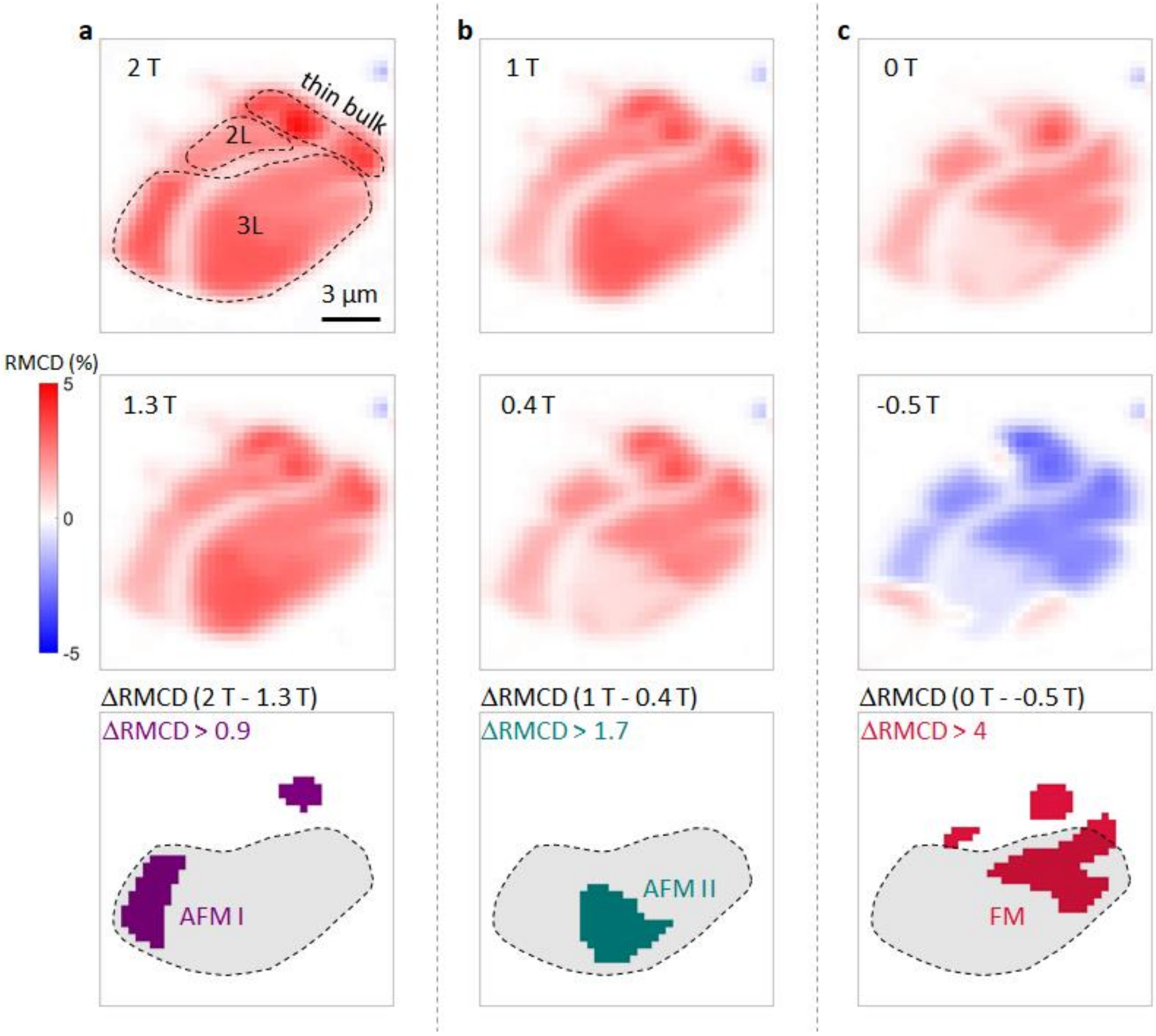

**Supplementary Fig. 5 | Procedure of constructing domains of different magnetic phases in the trilayer $CrI_3$.** The three magnetic phases have distinct critical fields for spin-flip transitions. We first show how to obtain domains of AFM I phase, which has the largest critical field for the spin-flip transition among the three magnetic phases. As magnetic field is swept down, in the region of 2 to 1.3 T, only AFM I phase has a spin-flip transition, with a change in RMCD signal of 1.3% (maintext, Fig. 4b). We therefore obtain full spatial maps of the RMCD signal at **a**, 2 T (top) and 1.3 T (middle). We then take the difference of these two maps and filter out the signal smaller than 0.9%. This results in the AFM I magnetic domains (bottom). As the magnetic field continuous to sweep down, in the region of 1 to 0.4 T, only AFM II magnetic phase has a spin-flip transition (maintext, Fig. 4c). The change of RMCD is about 2%. We therefore obtain full spatial maps of the RMCD signal at **b**, 1 T (top) and 0.4 T (middle). The difference of both maps yields area of AFM II magnetic domain (bottom). As magnetic field continues to decrease, we obtain full spatial maps of the RMCD signal at **c,** 0 T (top) and -0.5 T (bottom). For this field, the RMCD signal of all three phases change (see Figs. 4b-d in the main text). However, the ferromagnetic phase has the largest change of about 4.1% among the three phases. Therefore, the difference of the maps at 0 T and -0.5T with signal larger than 4% yields the area with ferromagnetic phase (bottom panel).